\documentclass{article}

\usepackage{arxiv}
\usepackage{amsmath,amsfonts,amsthm}
\usepackage[utf8]{inputenc} 
\usepackage[T1]{fontenc}    
\usepackage{hyperref}       
\usepackage{url}            
\usepackage{booktabs}       
\usepackage{amsfonts}       
\usepackage{nicefrac}       
\usepackage{microtype}      
\usepackage{lipsum}		
\usepackage{graphicx}
\usepackage{doi}
\usepackage{orcidlink}
\usepackage{siunitx}
\usepackage{subfig}
\usepackage{xcolor}
\usepackage{float}
\hypersetup{
colorlinks=true,
linkcolor=black,
citecolor=red
}
\usepackage{makecell}
\usepackage{multirow}
\usepackage{physics}
\usetikzlibrary{chains,scopes,positioning,backgrounds,shapes,fit,shadows,calc,arrows.meta,shapes.geometric}
\usetikzlibrary{quotes,angles}
\usetikzlibrary{calc}
\usetikzlibrary{decorations.pathreplacing,math}
\RequirePackage{amsmath,mathrsfs,amsfonts}
\usepackage{circuitikz}
\definecolor{myNewColorA}{RGB}{29,104,174}
\definecolor{myNewColorB}{RGB}{56,133,196}
\definecolor{myNewColorC}{RGB}{98,164,208}
\definecolor{myRed}{RGB}{216,56,58}
\definecolor{myGreen}{RGB}{84,179,69}
\definecolor{myGrey}{RGB}{145,147,147}
\definecolor{waterColor}{RGB}{38, 82, 140}
\definecolor{colaColor}{RGB}{60, 146, 166}
\definecolor{alcoholColor}{RGB}{242, 211, 153}
\definecolor{lemonColor}{RGB}{217, 171, 115}
\definecolor{oilColor}{RGB}{166, 118, 101}
\definecolor{pepsiColor}{RGB}{237, 178, 32}
\definecolor{spriteColor}{RGB}{183, 39, 62}
\definecolor{solutionColor}{RGB}{0, 114, 189}
\colorlet{Ecol}{orange!90!black}
\colorlet{EcolFL}{orange!80!black}
\colorlet{veccol}{green!45!black}
\colorlet{EFcol}{red!60!black}
\colorlet{pluscol}{red!60!black}
\colorlet{minuscol}{myNewColorA}
\usepackage{bm}
\def\systemname{\emph{WiField}}
\def\netname{\emph{Painter-Net}}
\title{The Field-based Model: A New Perspective on RF-based Material Sensing}

\author{Fei~Shang$^{\orcidlink{0000-0002-5495-8869}}$ \\
	University of Science and Technology of China\\
	\texttt{shf\_1998@outlook.com} \\
	\And
	Haocheng Jiang\\
	University of Science and Technology of China\\
	   \texttt{haochengjiang@mail.ustc.edu.cn} \\
	   \And
	   Panlong Yang$^*$$^{\orcidlink{0000-0003-1057-2793}}$ \\
		  Nanjing University of Information Science and Technology\\
		  \texttt{plyang@ustc.edu.cn} \\
	\And
	   Dawei Yan$^{\orcidlink{0000-0002-9848-3017}}$ \\
	   University of Science and Technology of China\\
	   \texttt{yandw@mail.ustc.edu.cn} \\
\And
	Haohua Du$^*$$^{\orcidlink{0000-0002-8492-3990}}$ \\
	Beihang University\\
	\texttt{duhaohua@buaa.edu.cn} \\
 \And
	Xiang-Yang Li$^{\orcidlink{0000-0002-6070-6625}}$ \\
	University of Science and Technology of China\\
	\texttt{xiangyangli@ustc.edu.cn} 
}

\renewcommand{\shorttitle}{\textit{arXiv} Template}

\hypersetup{
pdftitle={The Field-based Model: A New Perspective on RF-based Material Sensing},
pdfsubject={},
pdfauthor={Fei Shang, et al.},
pdfkeywords={RF sensing, IoT, material identical},
}

\begin{document}
\maketitle

\begin{abstract}
	This paper introduces the design and implementation of \systemname, a WiFi sensing system deployed on COTS devices that can simultaneously identify multiple wavelength-level targets placed flexibly.
	Unlike traditional RF sensing schemes that focus on specific targets and RF links, \systemname\ focuses on all media in the sensing area for the entire electric field.
	In this perspective, \systemname\ provides a unified framework to finely characterize the diffraction, scattering, and other effects of targets at different positions, materials, and numbers on signals.
	The combination of targets in different positions, numbers, and sizes is just a special case.
	\systemname\ proposed a scheme that utilizes phaseless data to complete the inverse mapping from electric field to material distribution, thereby achieving the simultaneous identification of multiple wavelength-level targets at any position and having the potential for deployment on a wide range of low-cost COTS devices.
	Our evaluation results show that it has an average identification accuracy of over 97\% for 1-3 targets (5 cm $\times$ 10 cm in size) with different materials randomly placed within a 1.05 m $\times$ 1.05 m area.
\end{abstract}


\section{Introduction}
Material identification system using radio frequency (RF) signals should ideally satisfies the
following three necessary requirements:

(1) {The flexibility of target positions.} After the system is deployed, it should be able to identify targets anywhere in domain, rather than only identifying targets placed in a specific position.

(2) {The arbitrariness of target quantity.} Since that there is usually more than one item in the sensing domain, it should be able to identify all items simultaneously, whether the target is one or more.

(3) {The practicality of target size}. It should not only be applicable to large-scale targets, but also need to be able to identify wavelength-level targets. This is particularly important for the sub-6G signal, as a large number of targets such as books and beverages are at this scale.

If the above three requirements are met, we can imagine that WiFi-based material
identification will have a real opportunity to provide the necessary static
environmental information for existing smart home, VR/AR and other systems,
give full play to the ubiquitous characteristics of RF
signals, and provide a useful supplement to existing sensing systems (such as
vision-based).


However, to the best of our knowledge, no RF based sensing system that
satisfies all three conditions exists. Tagscan~\cite{wang2017tagscan},
LiquID~\cite{dhekne2018liquid}, LiqRay~\cite{shang2022liqray},
WiMi~\cite{fengWiMiTargetMaterial2019}, etc., have achieved high-resolution
liquid identification based on sub-6G signals. However, {they only can identify the target placed in placed in a specific position for identification}.
FG-Liquid~\cite{liang2021fg} and
LiquImager~\cite{shangLiquImagerFinegrainedLiquid2024} have realized
position-invariant material identification, but {they can only identify
    a single target at one time}. Wi-Painter~\cite{yanWiPainterFinegrainedMaterial2023} has
achieved multi-target identification. However, {it only can identify the target whose size is much larger than the wavelength}, which is hard to work in real applications.


\begin{figure}
    \centering
    \includegraphics[width=.7\linewidth]{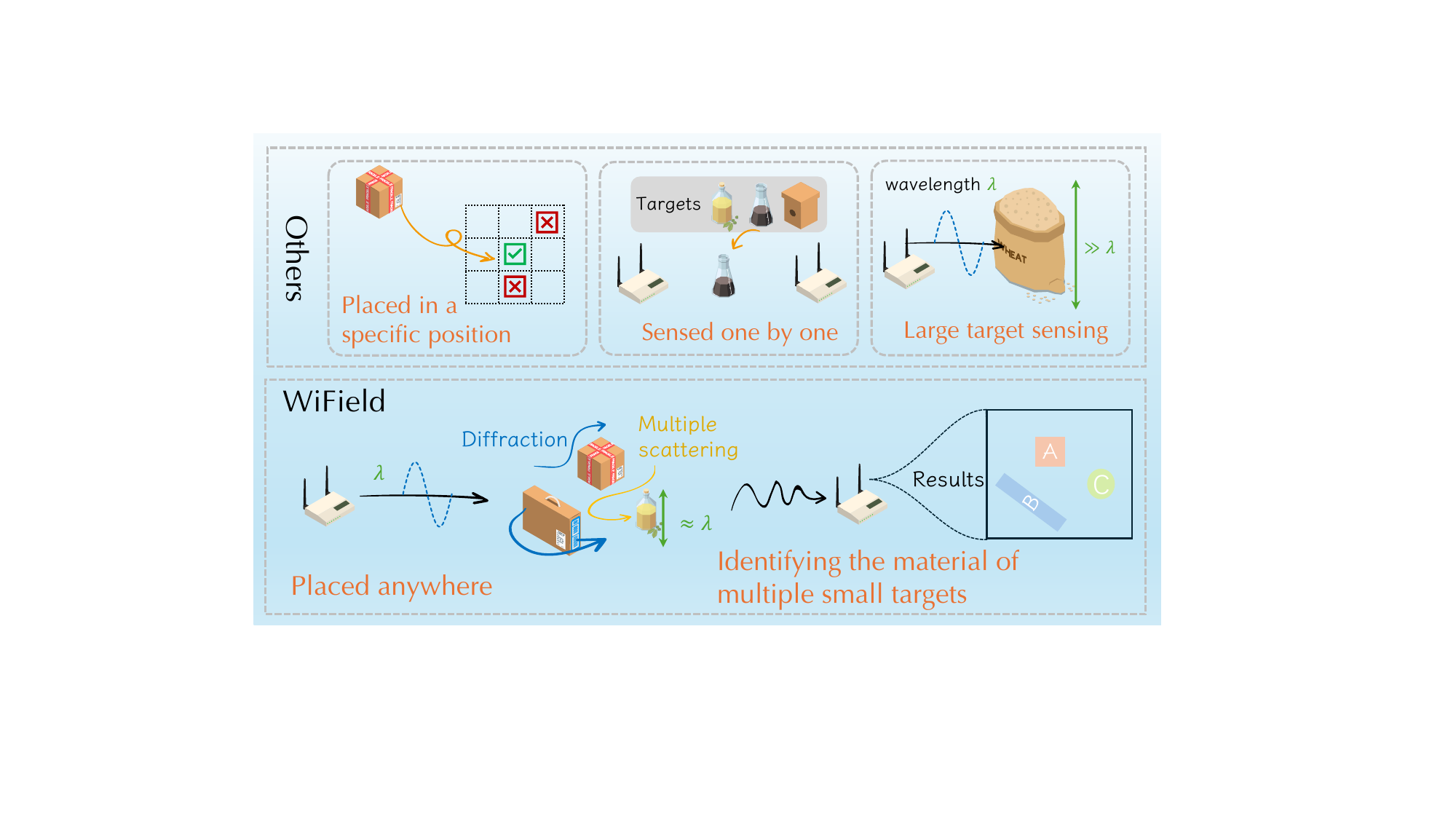}
    \caption{Characterizing the overall effect of different target combinations on the electric field in a unified framework.}
    \label{fig:challenges}
\end{figure}
In this paper, we propose \systemname, a material identification system that satisfies all the above necessary requirements.
We realize it with COTS WiFi Devices.
For three targets of different materials placed in a \SI{1.05}{m} $\times$ \SI{1.05}{m} domain with a size of \SI{5}{cm} $\times$ \SI{10}{cm}, \systemname\ can identify them simultaneously with an accuracy of more than 97\%.

{New perspective.} The opportunity for material identification comes from a basic observation: different materials have different effects on RF signals.
{Previous works mainly extract material information from a specific RF link} (usually equivalent to a {ray} starting from the transmitter and arriving at the receiver after reflection or transmission by the target).
Due to the need to carefully deal with reflection or transmission, such a system has a serious dependence on the target position.
In addition, due to the multiple reflections~\cite{IntroductionToElectrodynamics} of the signal between multiple targets, the applicability of the original system decreases sharply when the number of targets increases.
{Unlike previous work, we no longer focus on specific targets at specific locations and specific radio frequency links. 
Instead, we aim to describe the overall impact of the distribution of targets with different properties in space on the electromagnetic field.}
These properties can be various combinations of material, position, or quantity.
Within this framework, any specific combination is merely a special case.
This is not about special mappings between few point-to-point instances in the medium domain~\footnote{The set of combination of different material, position, or quantity.} and the signal domain, but rather an overall mapping from the entire medium domain to the signal domain. 
If we {construct a scheme to achieve the mapping from the received signal to the material distribution}, we can handle tasks involving the identification of attributes of multiple targets, multiple materials, and arbitrary locations simultaneously.
In this perspective, there is no substantial difference between a single target or multiple targets, or targets in different locations\footnote{There will be differences in the difficulty of solving.}.
As long as we make this scheme suitable for the perception of wavelength-level target objects, we have achieved our goal.
However, it is still difficult to implement field-based sensing on COTS WiFi devices.

(1) {How to describe the impact of different properties of a target on WiFi signals using a universal model?}
The material, position, and size of the target will all affect the signals, and their combination is uncertain.
In addition, the size of the target to be sensed is often close to the wavelength of the WiFi signal (for example, the wavelength of a WiFi signal with a frequency of \SI{2.4}{GHz} is \SI{12.5}{cm}), which brings obvious diffraction phenomenon.

(2) {How to recover the distribution of material from underdetermined signals?}
There are often hundreds or thousands of data to be estimated (for example, if we discretize the sensing domain into a $40 \times 40$ grids, we need to estimate 1600 parameters), and even if we use a MIMO WiFi system, there are only dozens even several independent WiFi links.

(3) {How to accurately identify with severe noise?}
We expect to recover the material of the target from the received signal with high resolution.
However, due to the presence of carrier frequency offset (CFO), etc., WiFi data contains severe phase noise, which further reduces the amount of information contained in the received signal.

{Our solutions are as follows.}
Firstly, we construct a field sensing model based on Maxwell's equations, which can more accurately describe the effects of multi-target diffraction and other factors on RF signals.
In order to use WiFi data with severe phase noise to estimate the distribution of material in the sensing domain, we do not directly solve the equation, but design an optimization method based on relatively stable amplitude data to find the solution that is ``most similar" to the real situation, and introduce the position information of the target as a regularization term to constrain the solution space.
Finally, we design \netname, using the non-linear characteristics of deep learning networks to improve the resolution.

{Contributions:} The main contributions in \systemname\ summarize as follows:
(1) We design \systemname, which can perform material identification using commercial WiFi devices. To the best of our knowledge, this is the first WiFi sensing system to achieve simultaneously identify multiple wavelength-level targets no matter where they are placed.

(2) We provide a new perspective for RF signal based sensing, which can describe the impact of different target combinations on the electric field within a universal framework, and has the potential for sensing in complex real-world scenarios.

(3) We propose a material identification scheme based on COTS WiFi devices and phaseless data, which verifies the feasibility of deploying this framework on low-cost commercial devices.


\section{Preliminaries}
\label{sec:pre}
\subsection{Antenna radiation field}
\label{sec:antenna}
WiFi devices (such as routers) usually have thin straight antennas.
We take thin antennas as an example to illustrate the radiation field.

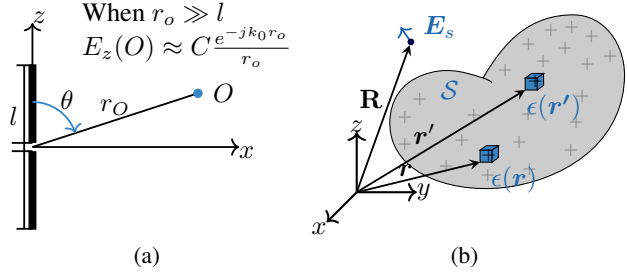
\begin{figure}[t]
    \centering
    \subfloat[]{\def \globalscale{0.55}
\def\linewidthEz{0.7pt}
\begin{tikzpicture}[y=1cm, x=1cm, yscale=\globalscale,xscale=\globalscale, inner sep=0pt, outer sep=0pt]
\coordinate (o) at (0,0);
\coordinate (a) at (0,2.5);
\coordinate (p) at (4,1.3);
\draw[->,line width=\linewidthEz] (0,0) -- (5,0) node[below right] {$x$};
\draw[line width=\linewidthEz] (-0.5,0.1) -- (0,0.1);
\draw[line width=\linewidthEz] (-0.5,-0.1) -- (0,-0.1);
\draw[->,line width=\linewidthEz] (0,0.1) -- (0,3) node[above right] {$z$};
\draw[line width=3pt] (0,0.1) -- (0,2);
\draw[line width=3pt] (0,-0.1) -- (0,-2);
\draw[line width=\linewidthEz] (0,0) -- node[above=2pt] {$r_O$} (p);
\pic["$\theta$", draw=myNewColorB, <-, angle eccentricity=1.3, angle radius=0.6cm,line width=\linewidthEz]
    {angle=p--o--a};
\filldraw[draw=myNewColorB,fill=myNewColorB] (p) circle (0.1);
\node[right=0.2 of p] {$O$};
\node[below right,text=black,align=left] at (1.2,3.5 ) {When $r_o \gg l$ \\ $E_z(O) \approx C \frac{e^{-jk_0 r_o}}{r_o}$};

\draw[|-|,line width=\linewidthEz] (-0.2,-2) --  (-0.2,2);
\node[left] at(-0.35,0.75){$l$};

\end{tikzpicture}\label{fig:thinAntenna}}
    \subfloat[]{\def\globalscale{0.4}
\def\linewidthEz{0.7pt}
\begin{tikzpicture}[y=1cm, x=1cm, yscale=\globalscale,xscale=\globalscale, inner sep=0pt, outer sep=0pt]

    \tikzstyle{vector} = [->,>=stealth,line width=\linewidthEz]
    \tikzstyle{charged}=[fill=gray!40!white]
\tikzstyle{darkcharged}=[fill=myNewColorB]
\tikzstyle{darkcharged1}=[fill=myNewColorB]
\tikzstyle{gauss surf}=[fill=myNewColorA]
\tikzstyle{gauss line}=[myNewColorB]
\tikzstyle{EField}=[->,line width=\linewidthEz,myNewColorA]
\tikzset{
  EFieldLine/.style={line width=\linewidthEz,EcolFL,decoration={markings,
                     mark=at position #1 with {\arrow{latex}}},
                     postaction={decorate}},
  EFieldLine/.default=0.5}
  \tikzstyle{measure}=[fill=white,midway,outer sep=2]
\tikzstyle{metal}=[top color=black!10,bottom color=black!20,middle color=black!5,shading angle=35]
\def\L{2.2}
\def\H{2.2}
\def\offset{2.0}
\def\W{0.30}
\def\Nx{5}
\def\Ny{5}
\def\H{3}
\def\W{5.8}
\def\h{0.35}
\def\w{0.42}
\coordinate (O)  at (-1.1,-0.4);
\coordinate (P)  at ( 0.7, 4.6);
\coordinate (Q)  at ( 3, 0.6);
\coordinate (Q1)  at (4.5, 3);
\coordinate (B)  at ( 4.1,-0.2);
\coordinate (L)  at ( 0.0, 1.9);
\coordinate (TL) at ( 1.1, 3.5);
\coordinate (TM) at ( 3.1, 3.4);
\coordinate (T)  at ( 6.1, 5.5);
\coordinate (R)  at ( 7.8, 3.5);
\coordinate (SL) at ( 4.5, 3.7);
\coordinate (SR) at ( 6.7, 3.3);

\draw[charged] 
    (B) to[out= 10,in=-90] (R)  to[out= 90,in= 0] (T) to[out=180,in= 60] (TM)
        to[out=162,in= 20] (TL) to[out=200,in=90] (L) to[out=-90,in=190] cycle;
  \draw[line width=0.4]
    (TM) to[out=162-180,in=144] ++(0.3,-0.15);

    \def\Nx{5}
    \def\xmin{0.5}
    \def\xmax{7.4}
    \foreach \i [evaluate={\x=\xmin+(\i-1)*(\xmax-\xmin)/\Nx; \y=0.18+0.12*(\x-3.4)^2;}] in {1,...,\Nx}{
      \node[text=gray] at (\x,\y) {\tiny{$+$}};
    }
    \def\Nx{4}
    \def\xmin{0.5}
    \def\xmax{8.1}
    \foreach \i [evaluate={\x=\xmin+(\i-1)*(\xmax-\xmin)/\Nx; \y=1.2+0.10*(\x-3.5)^2;}] in {1,...,\Nx}{
      \node[text=gray] at (\x,\y) {\tiny{$+$}};
    }
    \def\Nx{4}
    \def\xmin{1.0}
    \def\xmax{8.2}
    \foreach \i [evaluate={\x=\xmin+(\i-1)*(\xmax-\xmin)/\Nx; \y=2.1+0.09*(\x-3.6)^2;}] in {1,...,\Nx}{
      \node[text=gray] at (\x,\y) {\tiny{$+$}};
    }
    \def\Nx{6}
    \def\xmin{4.0}
    \def\xmax{8.1}
    \begin{scope}[rotate around={-6:(6.0,5.1)}]
      \foreach \i [evaluate={\x=\xmin+(\i-1)*(\xmax-\xmin)/\Nx; \y=5.1-0.41*(\x-6.0)^2;}] in {1,...,\Nx}{
        \node[text=gray] at (\x,\y) {\tiny{$+$}};
      }
    \end{scope}

    \def\Nx{3}
  \def\xmin{5.1}
  \def\xmax{7.3}
  \foreach \i [evaluate={\x=\xmin+(\i-1)*(\xmax-\xmin)/\Nx; \y=4.4-0.90*(\x-5.8)^2;}] in {1,...,\Nx}{
    \node[text=gray] at (\x,\y) {\tiny{$+$}};
  }
  \draw[->,thick] (O) --++ (-1.0,-1.0) node[below left] {$x$};
  \draw[->,thick] (O) --++ ( 2.0, 0.0) node[right] {$y$};
  \draw[->,thick] (O) --++ ( 0.0, 2.0) node[above] {$z$};
\draw[vector] (O) -- (Q) node[midway,above=0.1,xshift=-2,anchor=0] {$\bm{r}$};
\draw[darkcharged] (Q) |-++ (0.4,0.4) |- cycle;
  \draw[darkcharged] (Q) ++ (0,0.4) --++ (0.21,0.15) --++ (0.4,0) --++ (-0.21,-0.15) -- cycle;
  \draw[darkcharged] (Q) ++ (0.4,0) --++ (0.21,0.15) --++ (0,0.4)
                     node[midway,below=0.2,anchor=140,text=myNewColorA] {$\epsilon(\bm{r})$} --++ (-0.21,-0.15) -- cycle;
  \node[above right=0.02,text=black] at (Q) {\tiny{$+$}};
  \draw[darkcharged1] (Q1) |-++ (0.4,0.4) |- cycle;
  \draw[darkcharged1] (Q1) ++ (0,0.4) --++ (0.21,0.15) --++ (0.4,0) --++ (-0.21,-0.15) -- cycle;
  \draw[darkcharged1] (Q1) ++ (0.4,0) --++ (0.21,0.15) --++ (0,0.4)
                     node[midway,below=0.2,xshift=-2,anchor=140,text=myNewColorA] {$\epsilon(\bm{r'})$} --++ (-0.21,-0.15) -- cycle;
  \node[above right=0.05,text=black] at (Q1) {\tiny{$+$}};

  \draw[vector] (O) -- (Q1) node[midway,above=0.1,xshift=-2,anchor=0] {$\bm{r}'$};
  \draw[EField] (P) --++ (125:0.6) node[right=0.3] {$\bm{E}_s$};
  \node[fill=blue!30!black,circle,inner sep=0.9] (P') at (P) {};
  \draw[vector] (O) -- (P') node[midway,above left=0.2,anchor=-80] {$\vb{R}$};

  \node[text=myNewColorA] at (2,3) {$\mathcal{S}$};
  
\end{tikzpicture}\label{fig:scatter1}}
    \caption{Scattering field. (a) The scattering field intensity of the antenna is inversely proportional to the distance. (b) The scattering field of the medium is the result of scattered superposition and influence at various places.}
\end{figure}

In the far field, {the radiation field is a function of the distance from the antenna.}
As shown in Figure~\ref{fig:thinAntenna}, a thin straight antenna parallel to the $z$-axis, whose center is placed at the origin.
This antenna can only receive electric fields in the $z$ direction.
For a point $O$ in far field, whose distance from the origin is $r_o$,  we have $r_o \gg l$, where $l$ is antenna length.
The scattering field of the thin straight antenna at point $O$ is given by~\cite{krausAntennasAllApplications2002,orfanidis2002electromagnetic}
\begin{equation}
    E_z(O) =C \frac{e^{-jk_0 r_o}}{r_o},
    \label{eq:thinAntenna_new}
\end{equation}
where $C$ is a constant related to the antenna length, $k_0$ is the wavenumber, and $j^2=-1$.

\subsection{Scattering of EM waves by targets}
{Different materials have different complex permittivities}~\cite{IntroductionToElectrodynamics}, so the differences in the location, shape, and material of the target lead to different distributions of permittivities in the sensing domain, which cause varying degrees of scattering of EM waves.


{The scattering of EM waves by targets is determined by the material distribution.}
The EM waves propagating in space is completely determined by Maxwell's equations and boundary conditions, which is related to the distribution of permittivity.
Similar to Sec.~\ref{sec:antenna}, we consider the electric field in the $z$-axis direction.
As shown in Fig.~\ref{fig:scatter1}, the sensing domain $\mathcal{S}$ is marked in gray.
The relative permittivity at $\bm{r}$ is given by $\epsilon(\bm{r})$.
The electric field in $\mathcal{S}$ needs to satisfy~\cite{IntroductionToElectrodynamics,weiDeepLearningSchemesFullWave2019}:
\begin{equation}
    E_t(\mathbf{r}) = E_i(\mathbf{r}) + k_0^2 \int_\mathcal{S} G(\mathbf{r},\mathbf{r'}) J(\mathbf{r'}) d\mathbf{r'} \quad \text{for\ } \mathbf{r} \in \mathcal{S}
    \label{eq:maxwell1}
\end{equation}
where $E_t$ is the total field, $E_i$ is the incident field, $k_0$ is the wavenumber, $G$ is the Green's function, and $J$ is the equivalent current, which is given by $J(\bm{r})=[\epsilon(\bm{r})-1]E_t(\bm{r})$.
And the scattering field of $\mathcal{S}$ is given by
\begin{equation}
    E_s(\mathbf{R}) = k_0^2 \int_\mathcal{S} G(\mathbf{R},\mathbf{r'}) J(\mathbf{r'}) d\mathbf{r'} \quad \text{for\ } \mathbf{R} \notin  \mathcal{S}
    \label{eq:maxwell2}
\end{equation}
\section{Basic model}
\label{sec:model}
Since the complex permittivity is different, different materials scatter EM waves differently.
Unlike previous works that only focus on a few specific RF links, we consider to estimate the material distribution in the sensing domain as a whole.
{If we can infer the distribution of material in the sensing domain from the received signal, we have completed the multi-target sensing.}

{We directly construct a field model based on Maxwell's equations to explore the scheme of sensing multiple wavelength-level targets from the perspective of the sensing domain as a whole.}
We first establish a basic model, and then analyze the practical issues that need to be addressed during system construction.

\subsection{Field sensing model}
\label{sec:modelA}
{The field sensing model consists of two stages: (1) forward propagation and (2) inverse inference.}
The forward propagation phase is to calculate the scattered field based on the permittivity distribution of the sensing domain.
The reverse inference phase is to infer the permittivity distribution of the sensing domain based on the received signal.
In the forward propagation stage, we build a discretized multi-target scattering model based on Equ.~\ref{eq:maxwell1} and Equ.~\ref{eq:maxwell2}, to facilitate computation by the computer.
In the reverse inference stage, we employ an optimization scheme to infer the distribution of permittivity in the sensing domain.

\begin{figure}
	\centering
	\includegraphics[width=0.6\linewidth]{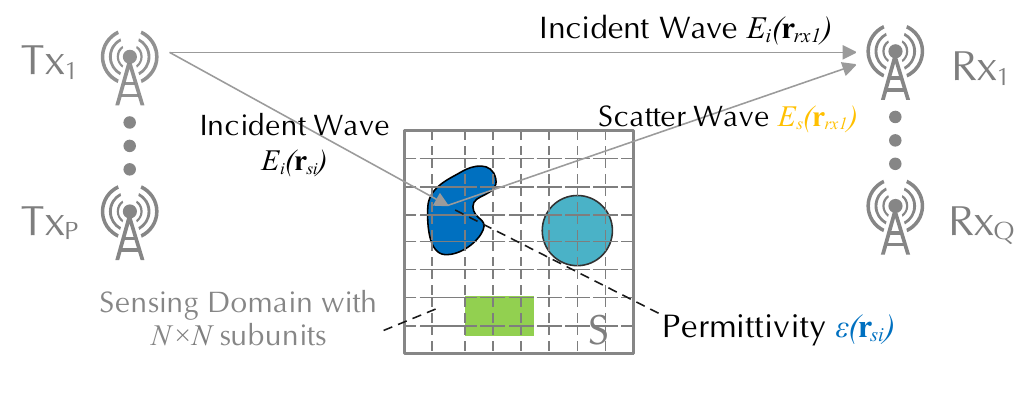}
	\caption{The total electric field at the receiving antenna is the superposition of the incident field and the scattered field from the sensing domain $\mathcal{S}$. The scattered field is related to the distribution of complex permittivity.}
    \label{fig:model}
\end{figure}
\begin{figure}
	\centering
	\subfloat[]{\includegraphics[width=0.3\linewidth]{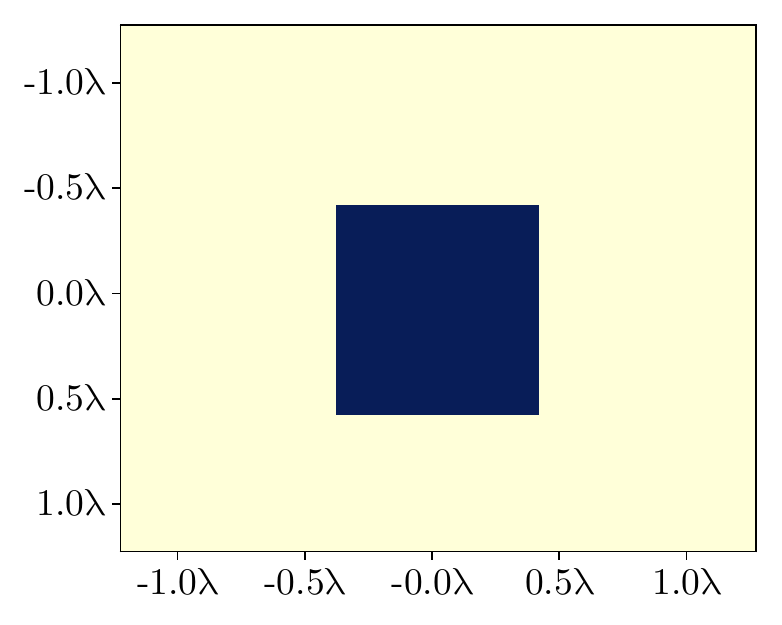}}
	\hspace{1em}
	\subfloat[]{\includegraphics[width=0.3\linewidth]{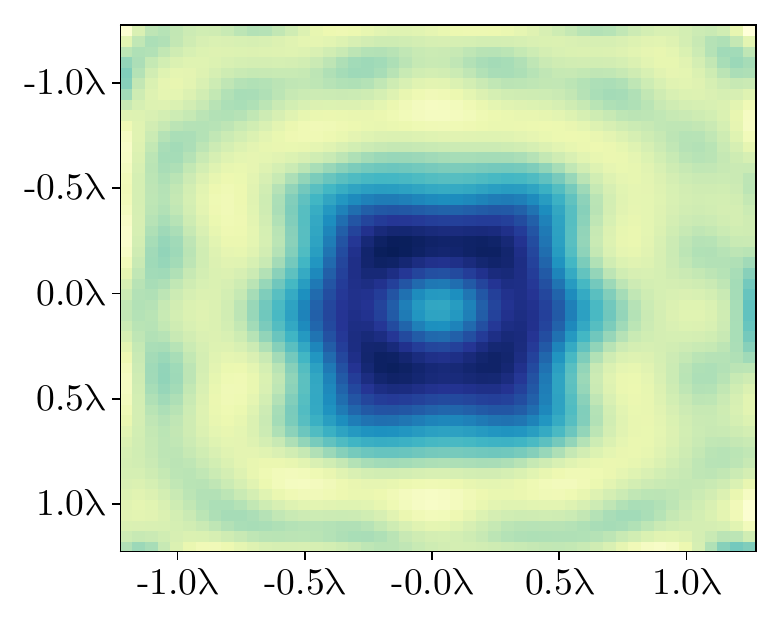}}
	\caption{The ground truth permittivity distribution (a) and the reconstructed permittivity distribution (b) using the Born approximation.}
	\label{fig:born}
\end{figure}

{(1) Forward propagation.}
We consider a two-dimensional top-down surface, which is perpendicular to a thin straight antenna.
{In this scenario, the electric field can be described using Equ.~\ref{eq:maxwell1} and Equ.~\ref{eq:maxwell2}.
For computational convenience, we discretized it~\cite{chen2018computational}.}
Additionally, we considered multiple targets and extended it to MIMO systems.

Specifically, as shown in Fig.~\ref{fig:model}, there are multiple targets of different materials and shapes in the sensing domain $\mathcal{S}$.
We divide the scattering domain into $N \times N$ sub-areas.
The position and relative permittivity of $i$-th sub-areas are $\mathbf{r}_{s_i}$ and $\epsilon_{s_i}$, respectively.
We assume there are $P$ transmitting antennas and $Q$ receiving antennas.
The position of the $p$-th transmitting antenna and the $q$-th receiving antenna are $\mathbf{r}_{tx_p}$ and $\mathbf{r}_{rx_q}$, respectively.
As a result, for the radiated field of the $p$-th transmitting antenna, Equ.~\ref{eq:maxwell1} and Equ.~\ref{eq:maxwell2} can be discretized as
\begin{equation}
	\left\{
	\begin{aligned}
	\mathbf{E}_t^p  &= \mathbf{E}_i^p + \mathbf{G}_{\mathcal{S}} \mathbf{J}^p \quad \text{for\ } \mathbf{r} \in \mathcal{S},\\
		\mathbf{E}_s^p  &= \mathbf{G}_O \mathbf{J}^p \quad \text{for\ } \mathbf{R} \notin  \mathcal{S}.
	\end{aligned}\right.
	\label{eq:maxwell3}
\end{equation}
The Green's coefficient matrix $\mathbf{G}_{\mathcal{S}}$ is $N \times N$ dimensions with $\mathbf{G}_{\mathcal{S}}[m,n] = k_0^2 A_{n} G(\mathbf{r}_{s_m},\mathbf{r}_{s_n})$ and $\mathbf{G}_O$ is $Q \times N^2$ dimensions with $\mathbf{G}_O[q,n] = k_0^2 A_{n} G(\mathbf{r}_{rx_q},\mathbf{r}_{s_n})$, where $A_{n}$ is the area of the $n$-th sub-area, $k_0$ is the wavenumber of the air, and $G(\mathbf{r},\mathbf{r'})$ is the 2-D free space Green's function.
The equivalent current density $\mathbf{J}^p$ is
\begin{equation}
	\mathbf{J}^p[i]= [\epsilon_{s_i}-1]E_t^p[i].
	\label{eq:J}
\end{equation}

\begin{figure}
	\centering
	\subfloat[]{\includegraphics[width=0.3\linewidth]{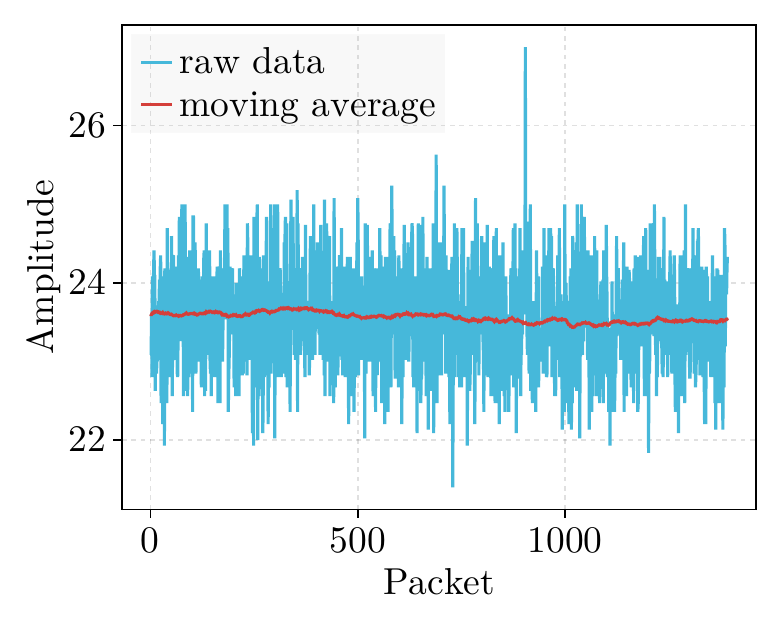}}
	\hspace{1em}
	\subfloat[]{\includegraphics[width=0.3\linewidth]{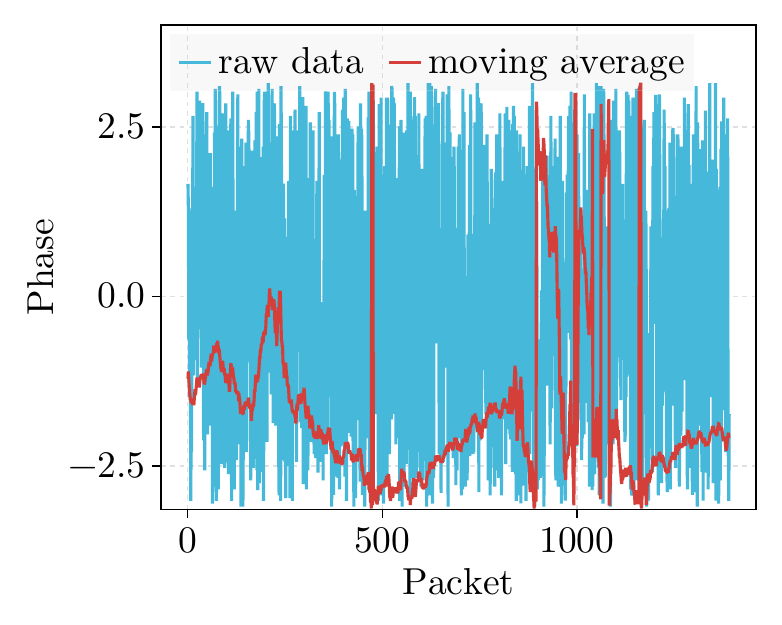}}
	\caption{The amplitude of WiFi signals is more stable than the phase after smoothing. (a) Amplitude. (b) Phase}
	\label{fig:signal}
\end{figure}


(2) {Inverse inference.}
In order to infer the distribution of permittivity using the scattering signals, an intuitive idea is to first calculate the equivalent current $\mathbf{J}$ with the scattering field $\mathbf{E}_s$, and then solve the Equ.~\ref{eq:maxwell3} and Equ.~\ref{eq:J} to get the permittivity $\epsilon$.
However, due to the filtering characteristics of the Green's function, the inverse problem of determining the induced current $\mathbf{J}$ is ill-conditioned, resulting in non-unique solutions for the $\mathbf{J}$.
{A typical approach is to simplify the scattering equation to a certain extent based on the properties of the scatterer, and then use an optimization scheme to obtain feasible solutions.}
Noting that many common materials used for reading, including glass, wood, rubber, etc., have weak conductivity and do not strongly scatter electromagnetic waves~\cite{PackquIDInpacketLiquid}, we use the Born approximation~\cite{devaneyInversescatteringTheoryRytov1981,slaneyLimitationsImagingFirstOrder1984} as the basis for solving inverse problems.

In the sensing domain $\mathcal{S}$, since the scattering field is weak compared to the incident field, we approximate that the total field $\mathbf{E}_t$ is equal to the incident field $\mathbf{E}_i$, so Equation~\ref{eq:J} becomes:
\begin{equation}
	\mathbf{J}^p[i] \approx  [\epsilon_{s_i}-1]E_i^p[i].
	\label{eq:J_born}
\end{equation}
Then, we solve an optimization problem using the least squares method to obtain the permittivity, with the objective function:
\begin{equation}
	\min_{\epsilon_{s_i}} \sum_{p=1}^P \Vert \mathbf{G}_{O} \mathbf{J}^p - \mathbf{E}_s^p \Vert^2 + \alpha \sum_{i=1}^{N\times N}\Vert \epsilon_{s_i}\Vert^2
	\label{eq:born}
\end{equation}
where $\alpha$ is the regularization parameter.

\subsection{Practical issues}
In practical applications, it is difficult to directly employ the sensing approach described above.
There are three practical issues that need to be addressed first.

{(1) The intermediate parameters required for equation solving are difficult to measure.}
The solutions to equations Equ.~\ref{eq:maxwell3} and Equ.~\ref{eq:born} depend on the scattered field at the receiving antenna and the incident field from the transmitting antenna to the sensing domain.
However, the received signal is the total field at the receiving antenna, which includes both the incident field from the transmitting antenna to the receiving antenna and the scattered field from the sensing domain to the receiving antenna.
The incident field at either the sensing domain or the receiving antenna is challenging to measure with receiving equipment, due to (a) the large number of sub-areas after discretization of the sensing domain, and (b) the presence of severe phase noise in the signal.

{(2) The phase of the received signal suffers from severe noise interference so that it is difficult to be used for complex problem solving.}
The solution approach introduced in Sec.~\ref{sec:model} relies on complex fields.
However, due to factors like carrier frequency offset (CFO), WiFi CSI data is often too noisy to be used directly.
Fig.~\ref{fig:signal} shows the amplitude and phase of the signals we collected.
After applying a mean filter with a sliding window width of 50 to 1400 sampling points, the amplitude data stabilized, but the phase data still suffered significant fluctuations.
Although there are some methods (such as the ratio method) that can suppress this noise, they typically depend on multi-antenna systems.
Noting the prevalence of single-antenna devices in daily life, we aim to design an sensing scheme compatible with single-antenna devices.


\section{System overview}
\label{sec:overview}

\begin{figure*}
    \centering
    \includegraphics[width=0.93\linewidth]{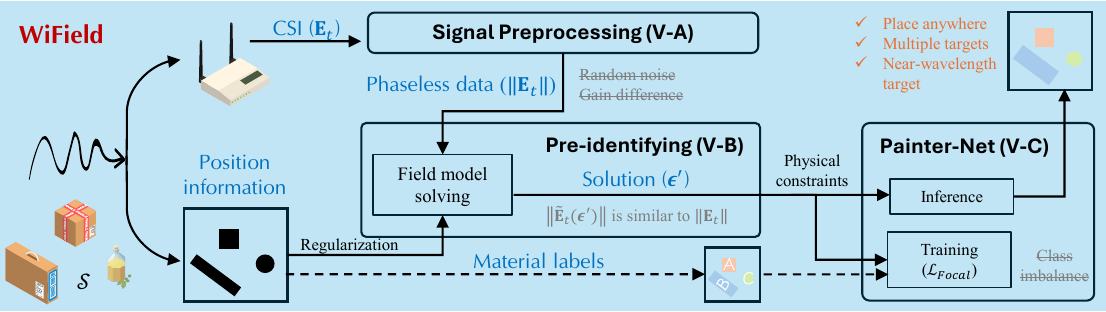}
    \caption{System overview.}
    \label{fig:overview}
\end{figure*}

Due to the fact that our field sensing model (Sec.~\ref{sec:model}) is directly based on Maxwell's equations, it accurately describes the impact of target position, diffraction, and mutual scattering between multiple targets on the received signal. This provides a theoretical basis for us to achieve multi-target, position-independent material identification based on WiFi signals.
However, the data required to solve the model are difficult to obtain directly by measurement.
Moreover, the Channel State Information (CSI) data collected by commercial Wi-Fi devices contains a lot of noise, especially in many cases where phase data is missing. This exacerbates the difficulty of multi-target material identification.
Towards position-independent material identification and imaging with CSI Data, We design \systemname.
The system schematic is shown in Fig.~\ref{fig:overview}.
Firstly, the CSI data obtained from the receiving antenna is used for signal preprocessing and pre-identifying, followed by image enhancement using a deep learning network.
Finally, we use the enhanced results for material identification and imaging.
Specifically, it consists of three parts:

(1) {Signal preprocessing}. We preprocess the collected data, including smoothing of amplitude data, outlier handling, and suppress the influence of device AGC gain.

(2) {Pre-identifying}. We estimate the intermediate variables required for solving the inverse process from the received signals, and a solution scheme suitable for phaseless data is designed to complete the preliminary estimation of permittivity using CSI data with severe phase noise.

(3) {\netname}. We design \netname, leveraging the non-linear characteristics of deep learning networks to further enhance material identification accuracy.

\section{System Design}
\label{sec:system_design}
The data we collected includes labeled data and RF data.
For each collection, we obtain the labeled data in a supervised manner, where we first generate an $N \times N$ matrix corresponding to the discretized sensing domain $\mathcal{S}$, and label each element with the corresponding material.
The RF data is the CSI value provided by the COTS WiFi device.
After obtaining the data, we first preprocess the RF data (Sec.~\ref{sec:incidence_field_estimation}) to suppress noise and interference from different device gains.
Noting that the phase of the signal contains a lot of noise, we design a inverse scheme (Sec.~\ref{sec:physical_parameters_estimation}) based on phaseless data and enhance the results using a deep learning network (Sec.~\ref{sec:deep_imaging_network}).

\subsection{Signal preprocessing}
\label{sec:incidence_field_estimation}
In order to use WiFi data for material identification in the field sensing model, we first need to deal with two problems: {data noise and gain difference}.
Since we expect to use amplitude data for subsequent tasks, since amplitude data is relatively stable, in the data noise part, {we only perform mean filtering and outlier removal on the original data.}
Since we use multiple independent transmitters and receivers, they may have different gain levels for signals.
{We suppress the influence of device gain differences through a single calibration.}

{Calibration scheme}.
We define the radiated field at position $\mathbf{r}$ by a transmitting antenna with a gain of $1$ as $E(\mathbf{r})$.
The scattered field and incident field excited by this radiated field are denoted by $E_{s}(\mathbf{r})$ and $E_{i}(\mathbf{r})$, respectively.
We will introduce our sensing model using an example of a transmitting antenna with a gain of $g_t$ and a receiving antenna with a gain of $g_r$.
As a result, when there are no objects present, the received signal $Y_{w/o}$ is equal to the incident field $\hat{E}^{inc}$, which is given by
\begin{equation}
    Y_{w/o} =\hat{E}_{i} = g_r g_t E_{i}(\mathbf{r}_{tr}),
    \label{eq:ywo}
\end{equation}
where $\mathbf{r}_{tr}$ is the position of the transmitting antenna.
When there are objects in the sensing domain, the received signal $Y_{w}$ is equal to the superposition of the incident field $\hat{E}_{i}$ and the scattered field $\hat{E}_{s}$.
Since the superposition of electric fields is a linear operation, we have
\begin{equation}
    \begin{aligned}
        Y_{w} &= \hat{E}_{i} + \hat{E}_{s} = g_r g_t \left[E_{i}(\mathbf{r}_{tr}) + E_{s}(\mathbf{r}_{tr})\right].
    \end{aligned}
\end{equation}
For common Wi-Fi antennas (thin straight antennas), based on Equ.~\ref{eq:thinAntenna_new}, the incident field $E_{i}$ can be expressed as
\begin{equation}
E_{i}(\mathbf{r}) =  \frac{e^{-jkr}}{r}
\label{eq:einc}
\end{equation}.

Based on these characteristics, we estimate the intermediate variables needed for solving the model.
The specific steps are as follows.

{STEP 1: Estimate the antenna gain $g_t g_r$.}
We first acquire the signal at the receiving antenna when no target is present in the space.
Then, we combine Equ.~\ref{eq:ywo} and Equ.~\ref{eq:einc} to estimate the $g_t g_r$, which is given by $g_t g_r = r||Y_{w/o}||$.

{STEP 2: Estimate the total field $E^{t}(\mathbf{r})$.}
Having get $g_t g_r$, we can estimate the scattered field $E^{t}(\mathbf{r})$ by $E^{t}(\mathbf{r}) = {Y_{w}}/{g_t g_r}$.

{STEP 3: Estimate the incident field in the sensing domain.}
Since the sensing domain is artificially divided by us, this implies that the location of each sub-area is known. 
Therefore, for the $i$-th sub-area (assuming its position is $\mathbf{r}_{ts_i}$), according to Equ.~\ref{eq:einc}, the incident field of the transmitting antenna in this sub-area is denoted as $E_{i}(\mathbf{r_{ts_i}}) =  {e^{-jkr_{ts_i}}}/{r_{ts_i}}$.

\begin{figure}
    \centering
	\subfloat[]{\includegraphics[width=0.3\linewidth]{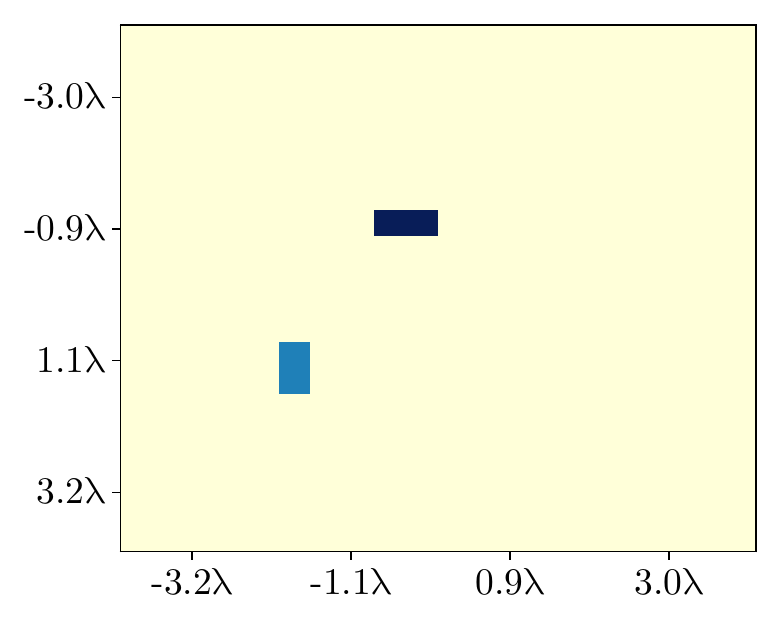}}
    \hspace{1em}
	\subfloat[]{\includegraphics[width=0.3\linewidth]{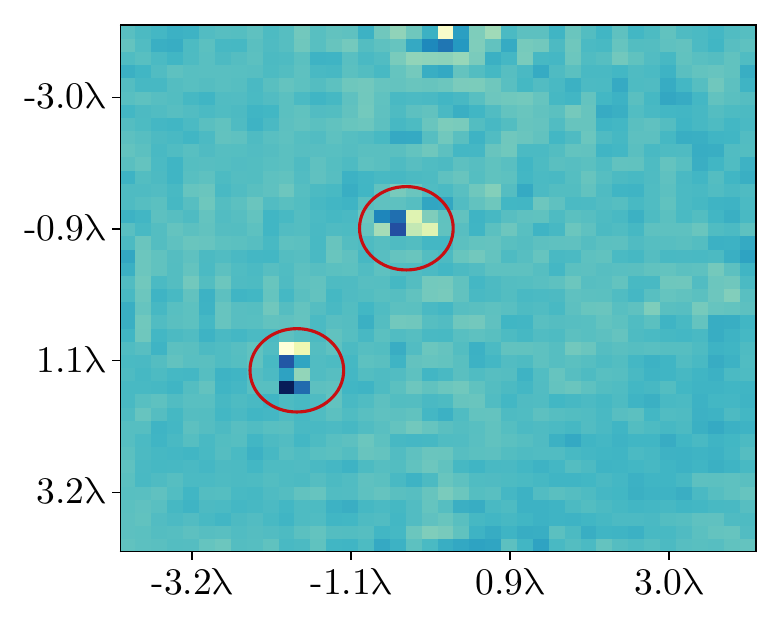}}
	\caption{The ground truth permittivity distribution (a) and the results (b) using pre-identifying.}
	\label{fig:born_phaseless}
\end{figure}
\begin{figure*}
    \centering
    \includegraphics[width=0.85\linewidth]{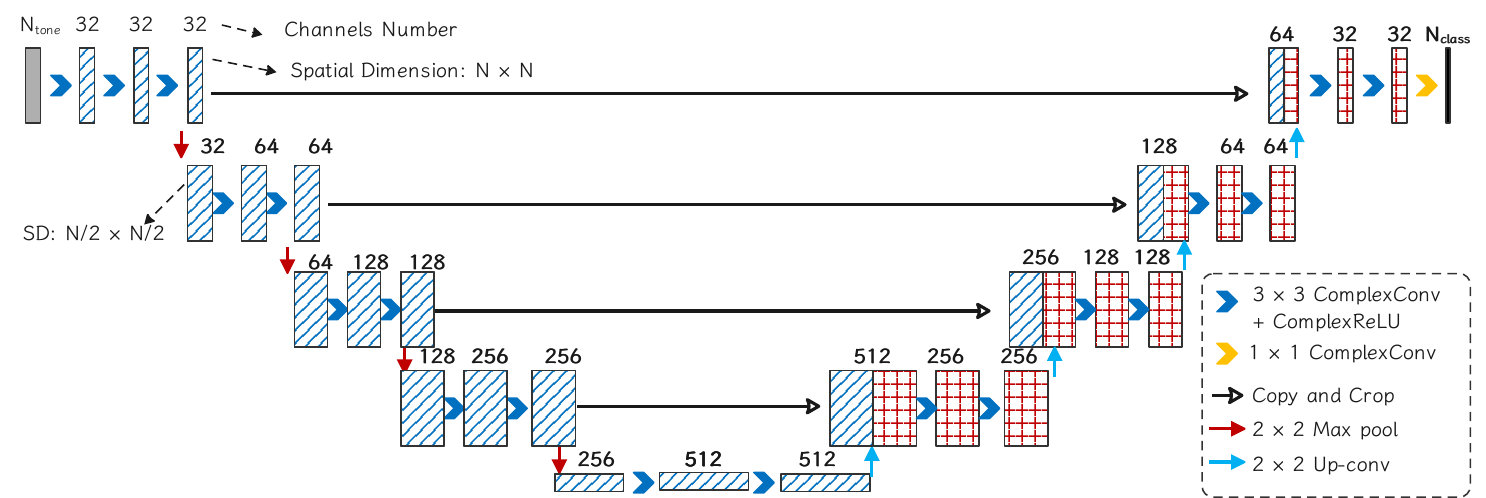}
    \caption{The structure of the network.}
    \label{fig:network}
\end{figure*}

\subsection{Pre-identifying}
\label{sec:physical_parameters_estimation}
If we can get the scattered field at the receiving antenna, then the complex permittivities of the sensing domain can be obtained according to the {inverse inference} introduced in Sec.~\ref{sec:model}.
Unfortunately, it is difficult to estimate the scattered field because we can hardly get the accurate phase of WiFi signal.
We therefore modified the inverse inference to estimate the distribution of permittivity solely from the {amplitude} of the WiFi signal.

Recall that the main idea of the Born approximation is to {obtain a set of permittivities such that the scattered field obtained by solving Maxwell's equations (Equ.~\ref{eq:maxwell3}) conditioned on it is most similar to the one measured, and a regularization term is used to ensure the convergence of the optimization problem.}
We modify the objective function (Equ.~\ref{eq:born}) based on this idea so that it can be applied to phaseless data.
Specifically, the objective function is given by
\begin{equation}
    \min_{\epsilon_{s_i}} \sum_{p=1}^P \underbrace{\left\Vert \frac{A^p}{\mathbb{E}[A^p]} - \frac{\Vert \hat{\mathbf{E}}_t^p \Vert^2}{\mathbb{E}[\Vert \hat{\mathbf{E}}_t^p \Vert^2]}\right\Vert^2}_{\text{constraint item}} + \underbrace{\alpha \mathcal{L}_{b}\left(f_{n}(\text{abs}.(\bm{\epsilon}),I\right)}_{\text{regularization item}},
	\label{eq:born_phaseless}
\end{equation}
where $\mathbb{E}[\cdot]$ denotes the expectation, $\hat{\mathbf{E}}_t^p$ is the assignment of CSI which is from the $p$-th transmitting antenna to the received antennas, $\alpha$ is the regularization parameter, $\mathcal{L}_{b}$ is the binary cross entropy (BCE) loss, $f_n(\cdot)$ represents the max-min normalization, $\bm{\epsilon}(i) = \epsilon_{s_i}$ is the vector of permittivities in the sensing domain $\mathcal{S}$,  $abs.(\cdot)$ represents magnitude by element, and $I$ is the the binary vector corresponding to the sensing domain $\mathcal{S}$ (where the target is located as 1, otherwise as 0).
$A$ is given by
\begin{equation}
    A = \Vert G_{O} \mathbf{J^p} + \mathbf{E}^p_{inc}\Vert^2
\end{equation}
where $\mathbf{J^p}$ is the induced current that is given by Equ.~\ref{eq:J_born}, $\mathbf{E}^p_{inc}$ is the incident field of the $p$-th transmitting antenna  in the received antennas.


In Equ.~\ref{eq:born_phaseless}, the constraint item is used to ensure that the estimated permittivities are consistent with the measured data, while the regularization item is used to ensure the convergence of the optimization problem.
In order to avoid interference from factors such as automatic gain control, making it difficult for us to obtain the actual value of the electric field in space, we have incorporated a mean normalization operation into the constraint term.
This means that we do not require the error between the estimated and measured values to be as small as possible. 
Instead, we compromise by requiring that the distribution of the estimated and measured values on each receiving antenna and each subcarrier is similar.

Fig.~\ref{fig:born_phaseless} (a) and Fig.~\ref{fig:born_phaseless}  (b) respectively display the ground truth and pre-identifying results.
We can roughly demonstrate that the target location is different from the surrounding air.
However, due to the existence of factors such as the non-linearity of the inverse problem, the results are very blurry. Therefore, we designed \netname\ to enhance these results.

\subsection{\netname}
\label{sec:deep_imaging_network}

Due to the filtering properties and non-linear factors of the Green's function~\cite{xiaoDualModuleNMMIEMMachine2020}, optimization algorithms often require a long time for iteration to obtain better solutions. 
Conversely, we attempt to view this problem from another perspective: we consider the complex permittivity estimated by the GOM algorithm after $k$ rounds of iteration as a distorted image, and the true value as an undistorted image, then use a neural network for image recovery. 
{Given the powerful non-linear fitting ability of neural networks, it will help improve the speed and accuracy of image recovery.}

This problem is a little bit similar to {image segmentation}: we want the network to output an image where targets of different materials (which have different complex permittivities) are assigned different colors.
In the field of image segmentation, the U-net~\cite{ronnebergerUNetConvolutionalNetworks2015} network has attracted much attention due to its simple structure and superior performance.

However, there are {three practical problems} to be solved before using the method for complex permittivity recovery:

(1) {Truth value is indeed}. For image restoration, we usually have a lossless image as the truth value. However, for permittivity recovery, it is difficult to obtain the complex permittivity of different materials in the perceptual region.

(2) {Categories are not balanced}. Unlike typical image segmentation tasks (such as VOC2012\cite{jingSelfsupervisedVisualFeature2019}), for multi-object sensing tasks, the side length of the sensing domain $\mathcal{S}$ often exceeds \SI{1}{m}, while the size of the perceived objects is very small (centimeter level).

For instance, there are several bottles of water placed on a desk. This leads to an imbalance in the categories of foreground and background.
We need to carefully handle this issue to accelerate network convergence.

{Details of \netname.}
We modify the basic Unet network to design \netname\ so that it is suitable for multi-target material identification scenarios. The details of the network are as follows.

\begin{figure*}
    \centering
    \includegraphics[width=0.95\linewidth]{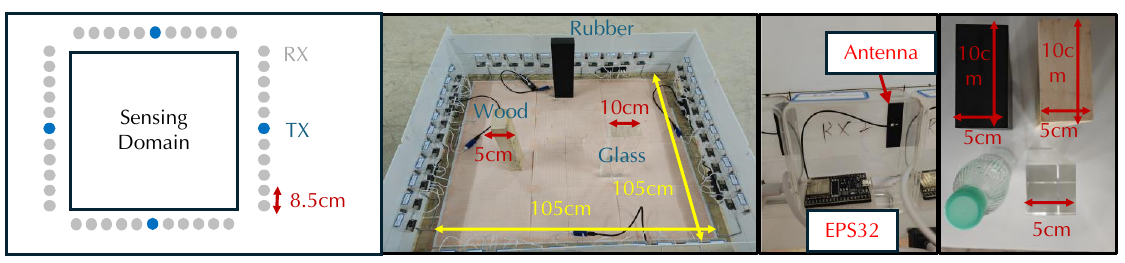}
    \caption{Experimental Setup. We use multiple inexpensive WiFi devices as transmitters and receivers, with a total cost of approximately 120\$. The size of the sensing domain $\mathcal{S}$ is \SI{1.05}{m} $\times$ \SI{1.05}{m}, and 1-3 targets of different materials are randomly placed within $\mathcal{S}$.The dimensions of the target are \SI{5}{cm} $\times$ \SI{10}{cm}, which is smaller than the signal wavelength.}
    \label{fig:exp}
\end{figure*}
(1) {Input.} Unlike conventional image tasks where the input is three-channel RGB data, the input for \netname\ is the  permittivity obtained through an inverse process.
Its dimensions are $[N_{tone},N,N]$, where $N_{tone}$ represents the number of effective subcarriers and $N$ is the number of grids contained in a row after discretizing the scattering area. 
Each element in the input is a complex number.

(2) {Structure.} The structure of the network is shown in Fig.~\ref{fig:network}. 
The overall structure of the network is similar to the typical Unet network, but in order to make it suitable for negative input values, we replace the main parts with complex networks, including convolutional layers, activation functions, pooling layers, and deconvolutional layers.

(3) {Output}.
To accomplish material identification, an intuitive way is to use the permittivities of the target domain as output. However, the measurement of permittivity often relies on dedicated instruments and has a large frequency difference, which makes the measurement of permittivity costly and complicated. Therefore, we use different labels to distinguish different materials. Specifically, if our goal is to distinguish $N_{class}-1$ kinds of materials, we use $N_{class}$ integers as labels, where $0$ represents air, and $1$ to $N_{class}-1$ represent the remaining other materials respectively.

(4) {Loss function.}
Due to the small size of the target to be identified, but it may appear in a wide range of locations, resulting in the target occupying a sparse position relative to the entire perception area. This causes an imbalance between background and foreground categories. Therefore, we assign different weights to labels of different categories for category balance. We use multi-target Focal loss for training, which is:

\begin{equation}
\mathcal{L}_{\text{Focal}} = -\sum_{i=1}^{N_{class}} \alpha_i (1-p_i)^\gamma \log(p_i),
\end{equation}
where $\alpha_i$ is the weight of the $i$-th category, $p_i$ is the probability of the $i$-th category, and $\gamma$ is the focusing parameter.

\section{Impalement}
\label{sec:imp}

\subsection{Hardware and software.}
We used 44 ESP32 modules for signal receiving and transmitting, with 4 of them for signal transmitting and 40 for signal receiving. Each ESP32 module has only one antenna and costs about 2.75 \$. Their total cost is about 120 \$, which is comparable to the cost of a common WiFi sensing system (e.g., the system built by ASUS AX86U).
The locations of these ESP32 modules are shown in Fig.~\ref{fig:exp}.
We used a PC for WiFi data collection, which is equipped with the system of Ubuntu 22.04. The sampling rate of the signal is \SI{100}{Hz}.

The code for signal preprocessing (Sec.~\ref{sec:incidence_field_estimation}) and pre-identifying (Sec.~\ref{sec:physical_parameters_estimation}) parts is implemented in Julia scripts \cite{datserisDrWatsonPerfectSidekick2020,dixitOptimizationJlUnified2023}, which are executed on a PC (AMD 7950X CPU, 64G memory).
Moreover, we build \netname\ with PyTorch that is one of the most popular frameworks for graph neural networks.
The training process is conducted on a server with an NVIDIA 4090, with Ubuntu 22.04 as the operating system.

\begin{table}[t]
    \centering
    \caption{The types of materials used in the experiment.}
    \label{tab:nums}
    \begin{tabular}{lc|lc}
        \toprule 
        \makecell{Combination\\of materials} & \makecell{Number of\\position\\combinations}  & \makecell{Combination\\of materials}& \makecell{Number of\\position\\ combinations}  \\ \hline
        Wood & 15  & Glass, Glass & 15  \\ 
        Rubber & 15  & Wood, Wood & 20  \\ 
        Glass & 15  & Rubber, Wood & 15  \\ 
        Rubber, Rubber & 15  & \makecell[l]{Rubber, Rubber,\\Rubber} & 15  \\ 
        Glass, Wood & 15  & \makecell[l]{Wood, Wood\\Wood} & 22  \\ 
        Glass, Rubber & 15  & \makecell[l]{Rubber, Wood,\\Glass} & 20  \\ 
        \bottomrule 
    \end{tabular}
\end{table}
\subsection{Dataset and details.}
\label{sec:dataset}
We collected data focusing on three types of solid materials, including wood (\SI{5}{cm} $\times$ \SI{10}{cm}), rubber (\SI{5}{cm} $\times$ \SI{10}{cm}), and glass (\SI{5}{cm} $\times$ \SI{5}{cm}), as shown in Fig.~\ref{fig:exp}.
ESP32 operates on Channel 11 (\SI{2.462}{GHz}), with a wavelength of approximately \SI{12.5}{cm}, which is larger than the dimensions of all materials.
The size of the perception area was (\SI{1.05}{m} $\times$ \SI{1.05}{m}).
We discretize it into a 40 $\times$ 40 grids for pre-identifying (Sec.~\ref{sec:physical_parameters_estimation}).
For each set of materials, we randomly placed them at different positions within the perception area. There were a total of 197 combinations of materials and positions, with specific numbers shown in Tab.~\ref{tab:nums}.
For each combination, we collected data continuously for one second and then calculated the mean amplitude of this second's data for subsequent processing, and repeated this process 20 times.
In addition, in order to avoid the data obtained in this way being too approximate, we artificially add Gaussian noise with a variance of 0.2 to the completed data before it is input to \netname\ (added to both the training set and the test set).
The temperature of all materials is at room temperature between 16 $^{\circ}$C to 20 $^{\circ}$C.
The accuracy results are calculated from the average of 5 times.

\begin{figure*}
    \centering
    \begin{minipage}{.57\linewidth}
        \subfloat[]{\includegraphics[width=0.4\linewidth]{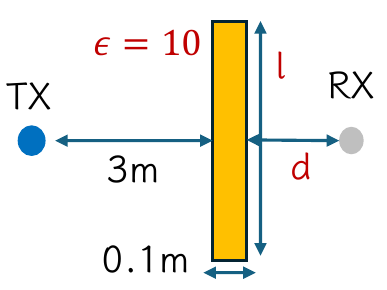}\label{fig:maxwell_vs_ray_exp}}
        \hfill
        \subfloat[]{\includegraphics[width=0.55\linewidth]{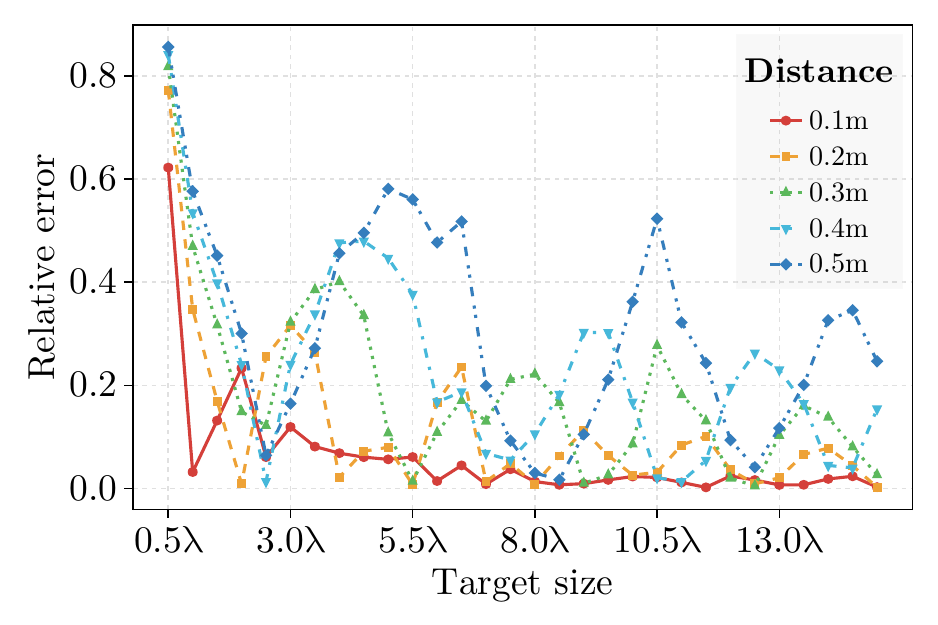}\label{fig:maxwell_vs_ray_res}}
        \caption{When the target size is relatively small, the relative error of the ray tracing model exceeds 80\%.}
    \end{minipage} 
\hfill
    \begin{minipage}{.34\linewidth}
        \centering
        \includegraphics[width=\linewidth]{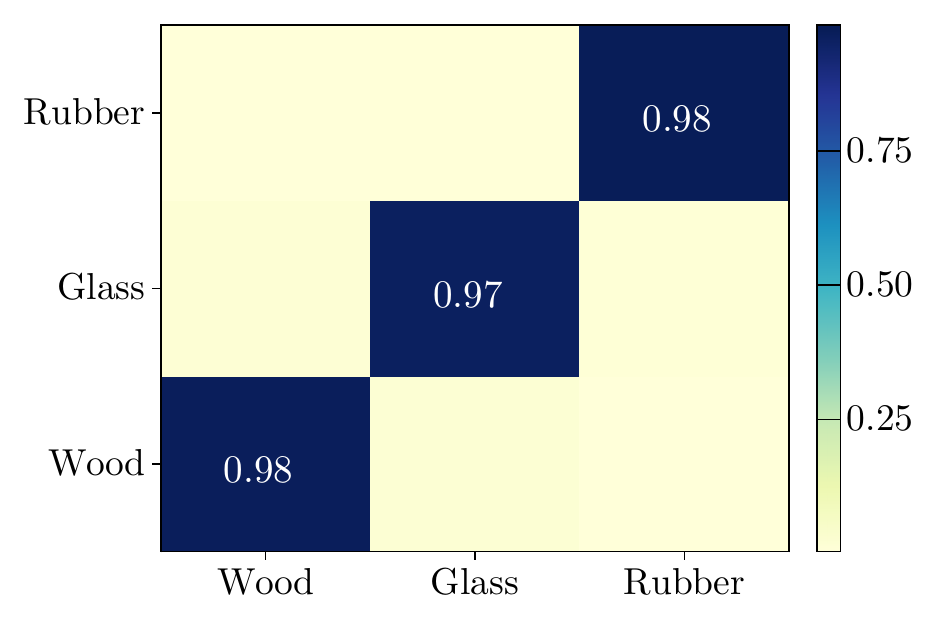}
        \caption{Recognition accuracy over 97\%.}
        \label{fig:baseline}
    \end{minipage}
\end{figure*}

For \netname, we set the number of input channels $N_{tone}$ to 30 (30 different subcarriers, with the same numbering as CSI TOOLS~\cite{Halperin_csitool}), and the number of output channels $N_{class}$ to 4.
The label values corresponding to air, wood, glass, and rubber are set to 0, 1, 2, and 3 respectively. 
The coefficient of Focal Loss is set to $0.005, 0.995,0.995,0.995$.
\section{EVALUATION}
\label{sec:eva}

\subsection{Why field model is necessary?}
\label{sec:why_field_model}

{The accuracy of the ray tracing model is not enough to complete multi-target with wavelength-level size sensing tasks. }
The influence of targets on radio frequency signals is mainly described by Maxwell's equations.
To facilitate calculation, we often use rays to approximate the propagation of radio frequency signals, which is a good approximation when the target size is much larger than the wavelength.
Unfortunately, many common targets to be perceived have sizes close to the wavelength of WiFi signals, which makes the signal suffer from severe diffraction and multi-level scattering interference, and it is difficult to be described by the ray tracing model.

Our simulation results indicate that when the target size is half a wavelength, the relative error between the signal amplitude obtained by the ray tracing model and the amplitude calculated by the Maxwell field model even exceeds {80\%.}
As shown in Fig.~\ref{fig:maxwell_vs_ray_exp}, we place one transmitter and one receiver on a horizontal line.
A target with a relative permittivity of $10$ is placed \SI{3}{m} to the right of the transmitter.
The width of the target is \SI{0.1}{m}, and its length $l$ varies from $0.5\lambda$ to $15\lambda$, where $\lambda$ is the wavelength.
We place the receiving antenna at $d$ to the right of the target, where $d$ varies from \SI{0.1}{m} to \SI{0.5}{m}.
The frequency of the signal is \SI{5}{GHz}.
Fig.~\ref{fig:maxwell_vs_ray_res} shows the relative error between the amplitude obtained by the ray tracing model and the results calculated by the field model.
We find that when $d$ is relatively large and $l$ is relatively small, the signal amplitude error is extremely large, implying that the radio frequency signal suffers severe diffraction effects.
In order to achieve high-precision identification of multiple wavelength-level targets, field model-based modeling is necessary.

\begin{figure}
    \centering
    \subfloat[Ground truth.]{\includegraphics[width=0.68\linewidth]{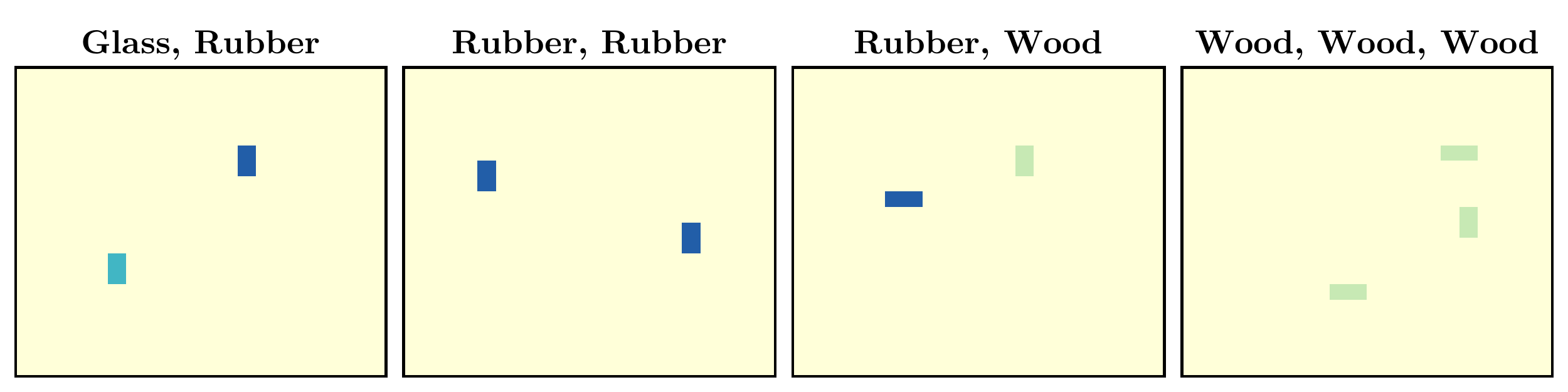}\label{fig:image_a}}

    \subfloat[The results of pre-idenfying.]{\includegraphics[width=0.68\linewidth]{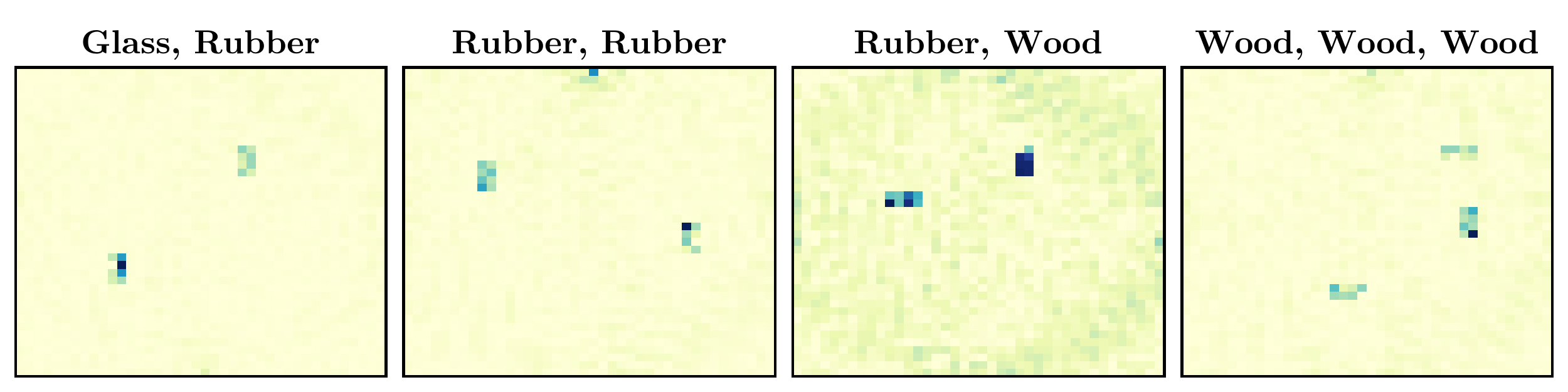}\label{fig:image_b}}

    \subfloat[The result of \netname.]{\includegraphics[width=0.68\linewidth]{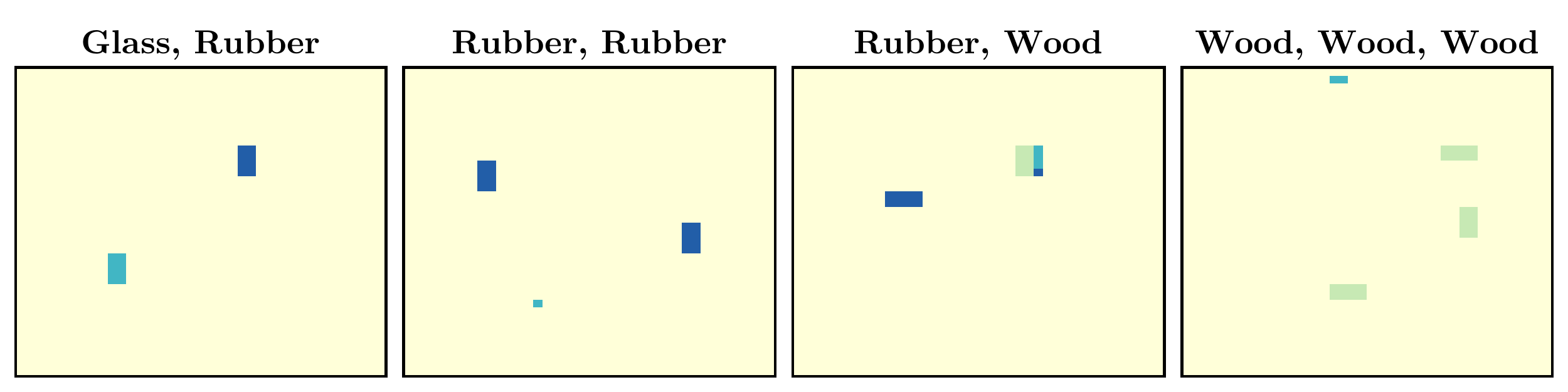}\label{fig:image_c}}
    \caption{Visualization of material recognition}
    \label{fig:image}
\end{figure}
\subsection{Experimental Results}

{Baseline.}
We first evaluate the \systemname's capability for material identification.
The dataset is constructed as described in Sec.~\ref{sec:dataset}.
Performance is assessed using five-fold cross-validation.
The output of \netname\ can take on four possible values (one or more of them), 0, 1, 2, 3, where 0 represents air and 1, 2, 3 represent three different solid materials.
The confusion matrix in Fig.~\ref{fig:baseline} illustrates the identification accuracy for different materials.
The results show that the average identification accuracy for the three materials exceeds 97\%.
Since the size of the targets to be identified does not exceed \SI{5}{cm} $\times$ \SI{10}{cm}, this means that both the length and width are smaller than the wavelength of the signal (approximately \SI{12.5}{cm} here).
This indicates that \systemname\ has the ability to identify multiple wavelength-level targets with high accuracy, which we attribute to our direct construction of a field model based on Maxwell's equations, allowing accurate characterization of the effect of multiple wavelength-level targets on RF signals (such as diffraction and multi-level scattering).

{Visualization.}
We randomly selected five cases for visual display, as shown in Fig.~\ref{fig:image}.
The combination and position of materials are different in different cases.
Fig.~\ref{fig:image_a} shows the ground truths of materials and positions in these five cases, with large areas of light yellow representing air (where there is no target) and other colors representing different materials, including glass, rubber, and wood.
Fig.~\ref{fig:image_b} shows the results of our pre-imaging using phase-free data (as described in Sec.~\ref{sec:physical_parameters_estimation}).
The results show that the distinction between target and air is relatively clear.
However, due to the existence of inverse nonlinear problems, the pre-imaging results are fuzzy and the consistency of materials is not good enough.
Fig.~\ref{fig:image_c} diagram shows the output of \netname, where the type and position of the material are very close to the ground truths, with only a very small number of pixels in error.
This shows that for different materials, different positions, different combinations of multiple wavelength-level size targets, \systemname\ has the ability to identify them.

{Ablation experiment.}
The simulation results of Sec.~\ref{sec:why_field_model} validate the necessity of the field model, and we design the ablation experiment to verify the role of \netname.
Specifically, we use different schemes to process the pre-imaging results, including k-nearest neighbor (KNN) classification, decision tree classification, support vector machine (SVM) classification, random forest classification, and \netname.
For machine learning algorithms such as KNN, we first binarize the truth value (air position is 0, target position is 1), and then multiply with the pre-imaging results to cluster the non-zero values into two categories, and then use the two clustering centers as material features for subsequent identification. The results are shown in Tab~\ref{tab:ablation}.
We found that the identification accuracy of the machine learning algorithm is difficult to exceed 40\%, which is far less than 97\% of \netname.
We believe that this is because the neural network has a strong nonlinear fitting ability and can better cope with the nonlinear factors existing in the reverse process.

\begin{table}[t]
    \centering
    \caption{Accuracy of Different Methods.}
    \label{tab:ablation}
    \begin{tabular}{lc}
        \toprule 
        Methods & Accurcay\\ \hline
        Pre-image + KNN & 27.0\%  \\
        Pre-image + Decision Tree & 31.3\%  \\
        Pre-image + SVM & 20.4\%  \\
        Pre-image + Random Forest & 31.3\%  \\ \hline
        Pre-image + \netname\ & 97.9\%  \\
        \bottomrule 
    \end{tabular}

\end{table}
\begin{table}[t]
    \centering
    \caption{Location generalization evaluation.}
    \label{tab:ablation2}
    \begin{tabular}{cc}
        \toprule 
        Proportion of training set & Accurcay\\ \hline
        60\% & 44.7\%  \\
        80\% & 71.1\%  \\
        90.0\% & 84.8\%  \\
        \bottomrule 
    \end{tabular}
\end{table}

{Position flexibility.}
Noting that in the baseline experiments, we put all samples together and then divide the data into training and test sets.
Since we conduct multiple independent acquisitions for the same combination of position and material during data collection, this is not sufficient to verify the system's position flexibility.
We re-divide the dataset, specifically, for the 197 combinations shown in Tab.~\ref{tab:nums}, we divide them into training and test sets at specific ratios.
If a combination is divided into the test set, we put all its samples into the test set, and vice versa.
Tab.~\ref{tab:ablation2} shows the identification performance of a when the proportion of the training set is different.
Under this dataset division method, the performance of a decreases, but for conventional training set proportions, such as 80\% or 90\%, \systemname\ can still maintain an accuracy rate of over 71\%.
And note that we divide the sensing domain $\mathcal{S}$ into 1600 small grids, for three different materials, if they are not limited by their positions, there are about $\sum_{i=1}^3\binom{3}{i}\binom{1600}{i}$ combinations in total, and we only collected signals for 197 cases, which is far less than 0.01\% of the total.
Moreover, as the diversity of samples in the training set increases, the systematic performance improves.
We believe that the field model-based identification scheme achieves good performance under sparse samples by limiting the solution space to a very small range through physical constraints, and \systemname\ has the ability to achieve position-independent multi-target material identification.
\section{Related work}
\label{sec:relatedwork}

\subsection{Wi-Fi based sensing}
The proliferation of WiFi devices has surged past 30 billion units, with over 13 billion currently in active use by consumers, facilitating widespread WiFi accessibility.
Academia and industry experts are zealously investigating the potential of WiFi sensing systems for diverse applications including tracking\cite{chi2022wi,qian2018widar2,xie2019md}, health surveillance\cite{yu2021wifi,zhangWicyclopsRoomscaleWiFi2024,wuWiFiCSIbasedDevicefree2022}, individual recognition~\cite{korany2019xmodal}, activity identification~\cite{adib2013see,gaoLiveTagSensingHumanobject2018}, imaging\cite{li2020wi,pallaproluWiffractNewFoundation2022}, among others.
Most pertinent to our study is the area of material recognition through WiFi technology.

\subsection{RF-based material identification method.}
In recent times, a plethora of outstanding cost-effective solutions utilizing wireless devices have emerged, greatly enhancing the pervasive deployment of material identification technologies.

{Relying on these devices, a multitude of exceptional projects have been put forward focusing on sensing capabilities.}
Based on the fact that different materials have varying effects on transmitted signals, LiqRay~\cite{shang2022liqray,shangContactlessFinegrainedLiquid2024}, PackquID~\cite{PackquIDInpacketLiquid}, WiMi\cite{fengWiMiTargetMaterial2019}, Wi-Fruit~\cite{liu2021wi} utilize sub-6G signals such as RFID and WiFi for fine-grained material identification or fruit ripeness monitoring.
Similarly, methods such as mSense~\cite{wuMSenseMobileMaterial2020}, LiqDetector~\cite{wangLiqDetectorEnablingContainerIndependent2023}, RFVibe~\cite{shanbhagContactlessMaterialIdentification2023}, and Tagscan\cite{wang2017tagscan} utilize millimeter wave or RFID devices to collect the reflection signals from targets in order to extract material information, achieving fine-grained monitoring of solid or liquid materials.
However, they can only identify the material at specific locations, which poses challenges for deployment.
FG-Liquid~\cite{liang2021fg}, Mmtaster~\cite{liangMmtasterMobileSystem2024}, and LiquImager~\cite{shangLiquImagerFinegrainedLiquid2024} explore the possibility of location-independent material identification.
However, they can only identify single targets, while in real scenarios, there are often multiple targets.
In addition, many excellent researches effort have attempted to use the polarization effect of materials on RF signals to extract features for material identification or imaging, among which Wi-Painter~\cite{yanWiPainterFinegrainedMaterial2023} has pioneered the integrated perception of multi-target imaging and identification.
However, they are often only applicable to targets much larger than the wavelength, and as the most widely existing signal, sub6G has a wavelength greater than 5cm, which greatly limits the broad applicability of the system.
Compared with these methods, \systemname\ can identify multiple targets with wavelength-level size, and can be deployed on COTS WiFi devices.

\section{Conclusion}
\label{conclusion}
In this paper, we propose \systemname, which can identification of multiple targets with wavelength-level size.
Owing to diffraction effects, the impact of containers sized in the centimeter range on the signal cannot be adequately represented through standard ray tracing techniques.
Commencing from Maxwell's equations as a foundation, we develop an electromagnetic scattering model for the electric field to elucidate how signals are influenced by centimeter-scale containers.
We design an optimization scheme to estimate material information from phaseless WiFi data, and then a deep learning network is designed to enhance the perception effect.
We deployed our system on COTS WiFi devices.
For wavelength-level targets of multiple different materials randomly placed in an area of \SI{105}{cm} $\times$ \SI{105}{cm}, \systemname\ can accurately identify them in it with more than $97\%$.
We believe \systemname\ can bring more possibilities for ubiquitous sensing.

\bibliographystyle{unsrt}
\bibliography{ref}  






\end{document}